\documentclass[12pt]{article}
\pdfoutput=1

\usepackage{graphicx}
\usepackage{amsmath}
\usepackage{amsfonts}
\usepackage{amssymb}
\usepackage[hidelinks]{hyperref}

\numberwithin{equation}{section}

\newcommand{\comment}[1]{}

\def\bsb{\boldsymbol}
\def\bea{\begin{eqnarray}}
\def\eea{\end{eqnarray}}
\def\be{\begin{equation}}
\def\ee{\end{equation}}

\def\d{\partial}
\def\ep{\epsilon}
\def\n{\bsb{\hat n}}
\def\P{\mathcal{P}}
\def\O{\mathcal{O}}
\def\cH{\mathcal{H}}
\def\x{{\bsb x}}
\def\v{\bsb v}
\def\q{{\bsb q}}
\def\k{{\bsb k}}
\def\p{{\bsb p}}

\newcommand{\mal}[1]{\mathcal #1}
\newcommand{\expect}[1]{\left\langle #1 \right\rangle}

\begin{document}

\vspace*{0.cm}
\begin{center}

{\LARGE\bf    CMB Anisotropies from a Gradient Mode}
\\[1.5 cm]
{\large Mehrdad Mirbabayi} and {\large Matias Zaldarriaga}
\\[0.7cm]

{\normalsize { \sl  School of Natural Sciences, Institute for Advanced Study, Princeton, NJ 08540}}\\
\vspace{.2cm}

\end{center}

\vspace{.8cm}

\hrule \vspace{0.3cm}
{\small  \noindent \textbf{Abstract} \\[0.3cm]
\noindent A linear gradient mode must have no observable dynamical effect on short distance physics. We confirm this by showing that if there was such a gradient mode extending across the whole observable Universe, it would not cause any hemispherical asymmetry in the power of CMB anisotropies, as long as Maldacena's consistency condition is satisfied. To study the effect of the long wavelength mode on short wavelength modes, we generalize the existing second order Sachs-Wolfe formula in the squeezed limit to include a gradient in the long mode and to account for the change in the location of the last scattering surface induced by this mode. Next, we consider effects that are of second order in the long mode. A gradient mode $\Phi = \q\cdot \x$ generated in Single-field inflation is shown to induce an observable quadrupole moment. For instance, in a matter-dominated model it is equal to $Q=5 (\q\cdot\x)^2 /18$. This quadrupole can be canceled by superposition of a quadratic perturbation. The result is shown to be a nonlinear extension of Weinberg's adiabatic modes: a long-wavelength physical mode which looks locally like a coordinate transformation. }

 \vspace{0.3cm}
\hrule

\section{Introduction and main results}

Hints of a north-south asymmetry in the CMB power spectrum \cite{Wmap,Planck,Akrami,Quartin} have led to several theoretical efforts for finding primordial explanations. One class of the models, originally proposed by Erickcek, Kamionkowski, and Carroll \cite{Erickcek}, postulates a long-wavelength perturbation that is a pure gradient in the observable part of the universe. If, in addition, there is a 3-point correlation function of the local type, the power-spectrum of short-wavelength modes is enhanced on one side of the sky and suppressed on the opposite side, leading to a dipolar power asymmetry. 

Erickcek et. al. first considered single field inflationary models and concluded that the induced power asymmetry would be unobservably small. The argument goes as follows. In momentum space, the momenta of the modes in a 3-point correlation function form a closed triangle. The local correlation corresponds to the squeezed limit where one of the modes has a much smaller momentum (longer wavelength) than the other two. In this limit one can approximate
\be
\label{fnl}
\frac{6}{5}f_{NL} \mal P_L \mal P_S=\expect{\zeta_L \zeta_S\zeta_S}'\simeq\expect{\zeta_L \expect{\zeta_S\zeta_S}_{\zeta_L}}'\simeq \mal P_L \frac{\d}{\d\zeta_L} \mal P_{S,\zeta_L},
\ee
where $\zeta$ is the conserved curvature perturbation, subscripts $L$ and $S$ indicate long- and short-wavelength. $\P_{L}\equiv \expect{\zeta_{L}^2}'$ is the power-spectrum of the long modes, and $\P_{S,\zeta_L}\equiv \expect{\zeta_S^2}_{\zeta_L}'$ is that of the short modes in the background of $\zeta_L$, and prime denotes the correlation function without the momentum conserving delta function.\comment{\footnote{In writing the left-hand side as a local non-Gaussianity we are ignoring scenarios with more singular than $1/k_L^3$ behavior.}} Suppose there is a long gradient mode which varies by $\Delta \zeta_L$ across the sky. According to \eqref{fnl} this modulates the power of the short modes and leads to the hemispherical power asymmetry 
\be
\label{A}
A\equiv \frac{\Delta \mal P_S}{\mal P_S}=\frac{1}{\mal P_S}\Delta\zeta_L \frac{\d}{\d\zeta_L} \mal P_{S,\zeta_L}=\frac{6}{5}f_{NL}\Delta \zeta_L.
\ee
There are various theoretical prejudices and observational constraints on $\Delta \zeta_L$ and $f_{NL}$. The wavelength of the long mode has to be very large to suppress its contribution to the CMB quadrupole and octopole. For perturbation theory to be valid the amplitude of the long mode was required in \cite{Erickcek} to be less than unity. These two requirements put an upper bound on $\Delta\zeta_L$, and the local non-Gaussianity $f_{NL}=5(1-n_s)/6 \sim 0.01$ of single field models \cite{Maldacena}, would be too small to produce observable power asymmetry. The authors then proposed a two-field (Curvaton) model in which $f_{NL}$, and hence $A$, can be larger. 

This scenario has been revisited and generalized by many authors since then (see e.g. \cite{Lyth}), but one part of the reasoning remained unchanged: the induced asymmetry in inflationary models that satisfy Maldacena's consistency condition $6f_{NL}/5=1-n_s$ is generally accepted to be small but {\em non-zero}. This seems to be in contrast with an argument about the physical part of the 3-point function \cite{Tanaka,Pajer}, which suggests that the induced asymmetry by a pure gradient must identically vanish in this case. The local non-Gaussianity dictated by the consistency condition is a consequence of the fact that a long-wavelength mode is locally equivalent to a coordinate redefinition, and so ref.s \cite{Tanaka,Pajer} argue no local observer should be able to measure it. In section \ref{null}, we will review this argument in the language of Weinberg's adiabatic modes \cite{Weinberg}. It will be shown that a linear gradient mode is equivalent to a coordinate transformation across the entire observable horizon, and thus, it has no dynamical effect on short-wavelength observables. In particular, its presence is inconsequential for the observed short-scale power, unless there is a correlation between the initial condition of short and long modes, corresponding to a violation of the consistency condition.

To make the general argument of section \ref{null} more explicit, and to improve our analytic understanding of one of the most important observables in cosmology, we will next perform a brute-force calculation of CMB anisotropies at second order. In section \ref{SW}, we review the derivation of the second order Sachs-Wolfe formula, and in section \ref{LS} specialize to the squeezed limit when one of the two modes is super-horizon and much longer than the other. The analysis generalizes the squeezed limit formula of \cite{Creminelli_squeezed} by keeping the gradient of the long wavelength mode, and confirms the expectation that a linear gradient mode does not effect CMB multipoles.

Finally, armed with the second order formalism we investigate another question of interest in section \ref{long}: What is the second order effect of a long gradient mode? The symmetry argument of section \ref{null} is no longer applicable since it treats the long mode linearly. We will find that there are observable effects including a finite contribution to the quadrupole moment. However, we will see that if the linear gradient initial condition is augmented by an appropriate quadratic piece, all observable effects vanish. This is next shown to be a second order extension of the Weinberg's adiabatic mode, which results in new concistency conditions on cosmological correlation functions.

The explicit analyses of sections \ref{LS} and \ref{long} are somewhat technical, so we first sketch the underlying ideas and summarize the main results.

\subsection{CMB anisotropies in the squeezed limit}

\begin{figure}
\centering
\includegraphics{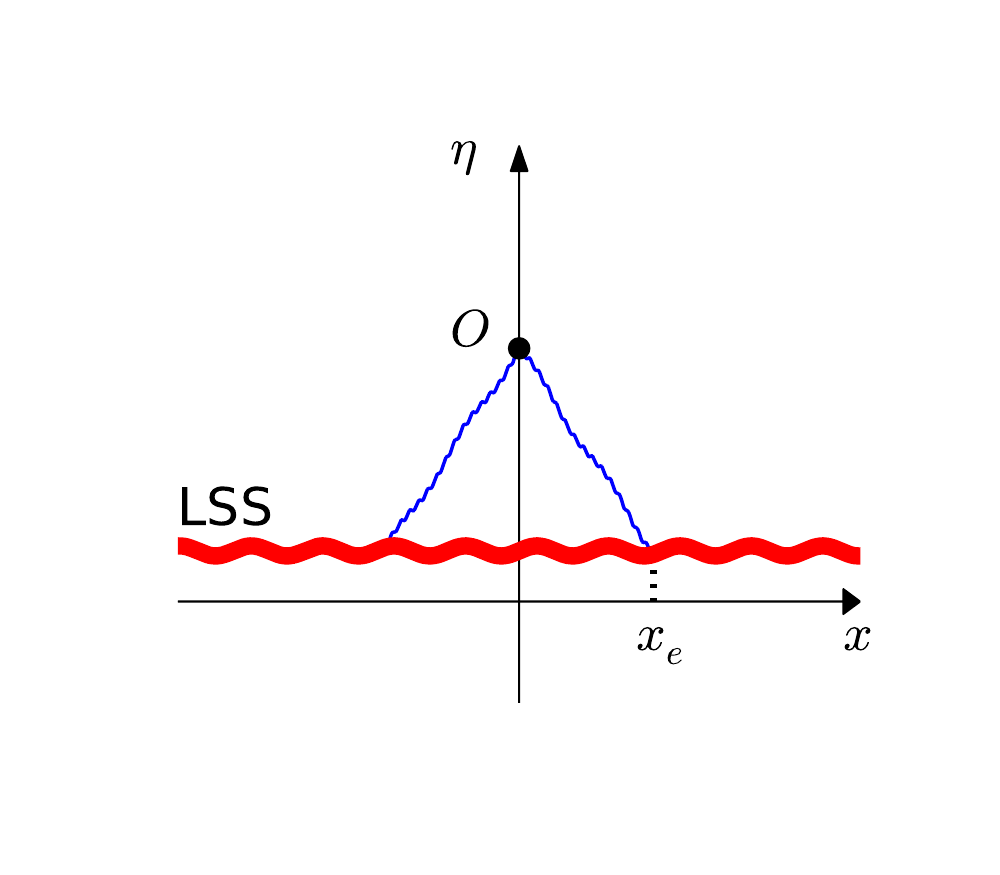} 
\caption{\small{At recombination a super-horizon perturbation is equivalent to a time-shift and a proportional coordinate rescaling. Afterwards it affects the propagation of CMB photons to the observer. If the mode enters the horizon in this period it leaves an observable imprint.}}\label{osclss}
\end{figure}

While analytic calculation of the second order effects generally involves complicated dynamics, considerable simplification occurs when one mode is super-horizon at the time of recombination and have much longer wavelength than the other. In this case, the effect of the long mode up until recombination is equivalent to a coordinate transformation (and possibly a variation of the initial condition), and the problem reduces to solving for the photon trajectory from the Last Scattering Surface (LSS) to the point of observation (see figure \ref{osclss}). This is the Sachs-Wolfe effect. 

In order to find the squeezed limit of CMB bispectrum, the leading 2nd order contribution to CMB anisotropies $\Theta(\n)$ from superposing a long and a short mode was analytically calculated in \cite{Creminelli_squeezed}. The result is incredibly simple
\be
\label{Th0}
\Theta(\bsb{\hat n})=\Theta_{S,\rm obs}+\Theta_{L,\rm obs}+\Theta_{L,\rm obs}\left(1+\frac{\d}{\d\ln\eta_e}-5\bsb{x}_e\cdot\nabla\right)\tilde\Theta_{S,\rm obs}
\ee
with
\be
\begin{split}
\Theta_{S,\rm obs}=&[\tilde\Theta+\tilde\Phi-\bsb{\hat n}\cdot \tilde{\bsb v}](\eta_e,\bsb x_e).\\[10pt]
\Theta_{L,\rm obs}=&\frac{1}{3}\Phi_L.
\end{split}
\ee
The quantities on the r.h.s. are evaluated at an early constant-time hypersurface $\eta_e$ (close to the recombination) and $\x_e$ is the emission point on that hypersurface. $\Theta,\Phi,\bsb v$ are respectively temperature anisotropy, Newtonian potential, and plasma velocity, and tilde corresponds to the value of linearly evolved short-wavelength quantities.

Though much longer than the horizon at recombination, the long mode in \eqref{Th0} can be much shorter than the present horizon. However, in the extreme case when the long mode is constant over the entire observable universe, it was shown by the authors that the formula predicts no observable effect of the long mode except for a homogeneous shift of the average temperature. However, the gradient of the long mode is ignored in this formula. Therefore, it is not possible to make the second consistency check, namely, unobservability of a linear gradient mode. We will generalize \eqref{Th0} to include gradient $\nabla \Phi_L$, and verify this expectation. However, one first needs to make the correct choice for the emission time $\eta_e$, as will be discussed next. 

\subsubsection{Recombination surface}

The Sachs-Wolfe formula can be used to express the observed temperature anisotropy in terms of quantities evaluated on -- and along the way to -- any earlier space-like hypersurface. It can, for instance, be a constant temperature $T=T_{\rm rec}$ (in an average sense), or a constant global time $\eta$ surface. The latter choice is commonly used in the derivation of Sachs-Wolfe formula, while the former choice seems to be preferred since temperature acts as a physical clock in cosmology. The difference between the two choices becomes relevant only at non-linear order, and one may worry that \eqref{Th0} expresses $\Theta(\bsb{\hat n})$ in terms of short-wavelength quantities at a surface whose temperature varies over the wavelength of the long mode. 

A related point is that in comparing the analytic formula \eqref{Th0} with the numerical results in the squeezed limit \cite{Creminelli_squeezed,Huang}, the time-derivative term on the r.h.s. was neglected, under the justification that it is suppressed by the ratio of the horizon sizes at recombination and today: $\eta_e/\eta_o \sim 0.001$. Contrarily, we expect
\be
\frac{\d}{\d\ln\eta_e}\Theta_{S,\rm obs} \sim c_s k_S\eta_e \Theta_{S,\rm obs}
\ee
to be large for the short modes that are well inside the horizon at recombination. As we will review in more detail, these time-derivatives result from the coordinate transformation $\eta_e \to \tilde \eta_e$ performed to remove the long mode. They will not be there in the first place if we express $\Theta(\bsb{\hat n})$ in terms of quantities at a constant temperature, or equivalently $\tilde \eta_e =\rm const.$, instead of quantities at a constant $\eta_e$.\footnote{We thank Paolo Creminelli for discussions on this point.}

Taking the gradient of the long mode into account, and choosing a constant-temperature hypersurface, we give the corrected generalization of \eqref{Th0} in section \ref{LS} [equation \eqref{theta1}], expressing CMB multiples to second order in terms of the linearly evolved fields. This is applicable when the long wavelength is much longer than the recombination horizon, but possibly shorter than the present horizon. Although because of rotational invariance the squeezed limit bispectrum (obtained by correlating the second order expression with two first order ones) cannot depend linearly on the gradient of the long mode, the result would in principle provide a new check of the second order numerical solutions besides the one performed in \cite{Creminelli_squeezed,Huang}. The formula passes the consistency check that when the long wavelength is much longer than present horizon (i.e. it is a pure gradient) there is no observable effect, thereby confirming the argument of section \ref{null}. 
\comment{
\be
\Theta(\bsb{\hat n})=\Theta_{S,\rm obs}+\Theta_{L,\rm obs}+\Theta_{L,\rm obs}\Theta_{S,\rm obs}+\Delta\bsb x_e\cdot \nabla \Theta_{S,\rm obs} - \Delta\bsb n \cdot \tilde{\bsb v}
\ee
where $\Delta \n$ and $\Delta\bsb x_e$ are defined respectively in \eqref{Dn} and \eqref{delT} in section \ref{LS}.}

\subsection{Second order effect of a gradient\label{adia2}}

While a gradient mode is equivalent to a coordinate transformation and hence unobservable at linear order, this is not necessarily the case at second order as for instance it produces a locally observable spatial curvature of order $(\nabla\zeta)^2$. At this order, a gradient mode contributes to CMB  monopole and quadrupole.\footnote{Although in practice it may be hard to break the degeneracy between CMB monopole and local effects, in principle they are distinguishable.} This arises firstly because of nonlinear terms in the Sachs-Wolfe formula, and secondly because unlike a uniform mode a gradient mode evolves in time at second order. This time-evolution has been calculated in a matter dominated universe in \cite{Matarrese,Bartolo,Creminelli_action}. For instance, given a generic initial condition $\zeta_0$, ref. \cite{Creminelli_action} found for $\zeta(t)$ 
\be
\label{z(t)}
\zeta(t) =  \zeta_0-\frac{1}{5 a^2 H^2}\d^{-2}\d_i\d_j(\d_i\zeta_0\d_j\zeta_0).
\ee
This solution is non-local. Thus for long-wavelength perturbations, the second order contribution to CMB observables would naively depend on how exactly they behave far outside our horizon. Obviously this cannot be the case: although various metric components are ambiguous, they should combine to give local and unambiguous expressions for CMB temperature. We will show that this is indeed the case. In particular for the initial condition $\Phi_0=-3\zeta_0/5 = \bsb q\cdot \bsb x+\phi_2$, where $\phi_2$ is a possible second order piece of $\O(q^2 x^2)$, the observed temperature is shown in section \ref{long} to take the following form
\be\label{Tcmb}
T(\bsb{\hat n})
=T_{\rm CMB}\Big[1-\frac{5}{36}q^2\eta_o^2+\frac{5}{18}(\bsb q\cdot \bsb x)^2+\frac{1}{3}\phi_2\Big].
\ee
Here $\eta_o$ is the observation time and $\x\simeq \n \eta_o$ is a point on the LSS. As expected, in the absence of $\phi_2$ there is a finite contribution of the gradient mode to the monopole and quadrupole moments, which puts an observational constraint on the variation of $\Phi$ across the sky $\Delta\Phi \sim q \eta_o$. 

As for the non-Gaussian correction $\phi_2$, it is natural to take it to be negligible if the gradient mode has been generated during a phase of minimal slow-roll inflation. The field $\zeta$ is almost free in this model, with interactions suppressed by slow-roll parameters \cite{Maldacena}, so we do not expect an order one non-Gaussianity to be produced by interactions. (In single-field models with derivative interactions too this sort of non-Gaussianity does not seem to arise.)

However, suppose the linear gradient had been superposed by
\be
\phi_2= \frac{5}{12} {q}^2 x^2-\frac{5}{6}(\q\cdot\x)^2 .
\ee
This would cancel the quadratic terms in \eqref{Tcmb}. We will show in section \ref{2nd} that this is a nonlinear generalization of Weinberg's adiabatic mode. That is, a physical mode which is locally equivalent to a coordinate transformation and hence unobservable. Projecting cosmological correlations on this second order adiabatic mode leads to new (double soft) consistency conditions which will be studied elsewhere \cite{double_soft,Joyce}.

\section{Unobservable non-Gaussianity\label{null}}

In this section, we present general symmetry arguments suggesting that a super-horizon long wavelength mode does not affect the short distance physics up to order $(\lambda_S/\lambda_L)^2$, where $\lambda_{S,L}$ are the short and long wavelengths. 

The calculation of the 3-point non-gaussianity of the single-field slow-roll inflation was first done by Maldacena \cite{Maldacena} in the comoving gauge. Using the Arnowitt-Deser-Misner parametrization the line element is written as
\be
ds^2 = -N^2 dt^2 + h_{ij}(dx^i+N^i dt)(dx^j+N^j dt),
\ee
and the spatial metric can be decomposed as
\be
\label{h}
h_{ij}=a^2 e^{2\zeta}\left(e^\gamma\right)_{ij},\qquad \gamma_{ii}=0.
\ee
The gauge is completely fixed by requiring $\d_i\gamma_{ij}=0$ and the inflaton field be unperturbed. Based on the form of \eqref{h}, Maldacena argued for a consistency check. A long-wavelength mode $\zeta_L$ is locally equivalent to a rescaling of coordinates $\tilde {\bsb x} = (1+\zeta_L)\bsb x$. Therefore, the squeezed limit of the 3-point function \eqref{fnl} can be calculated by replacing the 2-point function of short modes in the presence of the long mode, $\expect{\zeta_S^2}_{\zeta_L}$, by $\expect{\zeta_S^2}$ in the absence of $\zeta_L$ but evaluated at rescaled coordinates. Transforming to momentum space we get (see \cite{Creminelli_sct})
\be
\begin{split}
\expect{\zeta(\bsb x) \zeta(\bsb 0)}_{\zeta_L}&=\expect{\zeta(\tilde{\bsb x}) \zeta(\bsb 0)}=\int \frac{d^3\bsb k}{(2\pi)^3} \mal P(k) e^{i\bsb k\cdot (1+\zeta_L)\bsb x}\\[10pt]
&\simeq \expect{\zeta(\bsb x) \zeta(\bsb 0)}-\zeta_L \int \frac{d^3\bsb k}{(2\pi)^3} \mal P(k) e^{i\bsb k\cdot \bsb x} \frac{\d}{\d \ln k} \left[\ln (k^3 \mal P(k))\right] \nonumber\\[10pt]
&= \expect{\zeta(\bsb x) \zeta(\bsb 0)}-(n_s-1)\zeta_L\expect{\zeta(\bsb x) \zeta(\bsb 0)},\nonumber
\end{split}
\ee
giving $f_{NL} =5(1-n_s)/6$.\footnote{Conventionally, $f_{NL}$ is defined in terms of the Newtonian potential in a matter dominated universe $\Phi = g - f_{NL} g^2$ where $g$ is a Gaussian variable, and $\zeta = -5\Phi/3$.} 

To relate this consistency condition to the late time observables and see possible sources of violation, let us review the argument in the language of Weinberg's adiabatic modes \cite{Weinberg}. We proceed following Weinberg to find long-wavelength adiabatic solutions. First, we fix the conformal Newtonian gauge where the linearized metric looks like 
\be
ds^2 = a^2(\eta)[-(1+2\Phi)d\eta^2 +[(1-2\Psi)\delta_{ij}+\gamma_{ij}]dx^i dx^j],\qquad \gamma_{ii}=0.
\ee
This completely fixes the reparametrization freedom at non-zero momentum. However there are still homogeneous transformations which preserve the gauge condition:
\be
\label{diff0}
\eta\to \eta+\ep(\eta),\quad x^i\to (\delta^i_j+\omega^i_j) x^j, \quad \omega^i_j=\rm{const.}.
\ee
They generate a family of homogeneous solutions of the equations of motion
\be
\label{adia0}
\Phi = \ep'+\cH\ep,\quad \Psi =-\frac{1}{3}\omega^i_i -\cH\ep,\quad \gamma_{ij} = \omega_{ij}+\omega_{ji} - \frac{2}{3}\delta^i_j \omega^k_k
\ee
where indices are lowered by $\delta_{ij}$, $\cH=a'/a$ and prime denotes $d/d\eta$. A subfamily of these solutions can be shown to be extendible to non-zero momentum and therefore are physical. Those are the adiabatic modes. For this, one needs to ensure that the equations of motion are not accidentally satisfied because of an overall spatial derivative acting on the homogeneous fields \eqref{adia0}. For scalar modes and in the absence of anisotropic stress, this leads to the condition
\be
\label{PhiPsi}
\Phi=\Psi.
\ee
The solution is 
\be
\label{adia}
\omega^i_i=3 C_1,\quad \ep(\eta)=-\frac{C_1}{a^2(\eta)}\int^\eta a^2(\eta')d\eta' + \frac{C_2}{a^2(\eta)},
\ee
with $\{C_1,C_2\}$ constants. In cosmology we are mostly interested in the non-decaying mode, for which $\Phi,\Psi$ are given by
\be\label{diff13}
\Phi=\Psi =C_1\left(\frac{\cH}{a^2}\int a^2 d\eta-1\right),
\ee
which corresponds to the constant $\zeta = C_1$ in the comoving gauge. 

This argument, first of all, proves the existence of an adiabatic mode that remains conserved at super-horizon scales. Moreover, one can turn around the argument and see that once a cosmological perturbation is frozen into an adiabatic mode, its effect on short scale physics is just a coordinate transformation. Different points with different values of $\zeta_L$ experience the same history unless they start from different initial conditions. However, if at the time when the long mode freezes into an adiabatic mode the short modes are well inside the horizon and in vacuum (as in the case of attractor single-field inflationary models) the initial condition would also be the same. Thus, the only difference between points with different values of $\zeta_L$ is a relative shift in their history which is not locally observable \cite{Flauger}. Consequently, the squeezed limit non-gaussianity $6f_{NL}/5=1-n_s$ that follows from this relative rescaling does not lead to any measurable correlation between the short-distance physics and the long-wavelength mode. On the contrary, in multifield models with large local non-Gaussianity the adiabatic modes form at a later time when all $k$-modes are super-horizon and already excited. Hence, observable short-long correlations can be (and generically will be) generated.

What about correlation functions of CMB anisotropies? Since here we are observing photons coming from the surface of last scattering -- a distance $D$ away -- one would expect the long modes whose wavelength $\lambda_L$ are of order or less than $D$ correlate with short-wavelength modes with $\lambda_S\ll \lambda_L$. This is because for $\lambda_L<D$ the long mode is not equivalent to a coordinate transformation in the entire patch of size $D$, and it has a physical effect on the propagation of photons over this distance. On the other hand, a uniform mode with $\lambda_L\gg D$ must lead to no observable effect since it is equivalent to a coordinate transformation in the whole region of size $D$.

In this paper we are interested in a gradient mode about which the original Weinberg's argument does not have any immediate implication. The generalization to this case was given in \cite{Creminelli_sct,Hinterbichler_adia}. The zero-momentum coordinate transformation that can locally approximate a physical gradient mode $\zeta = \bsb q\cdot \bsb x$ is given by 
\be
\label{diff12}
\eta\to \eta+\ep(\eta,\bsb x),\qquad \bsb x\to (1+\bsb q\cdot \bsb x)\bsb x-\frac{1}{2}x^2 \bsb q -f(\eta)\bsb q , \qquad 
\ep =- f' \bsb q\cdot \bsb x,
\ee
where the physicality condition \eqref{PhiPsi} implies
\be
f'=\frac{1}{a^2}\int a^2 d\eta,
\ee
and the Newtonian potentials are now given by $\Phi =\Psi = (\cH f' -1)\bsb q\cdot \bsb x $.\footnote{The set of gauge preserving zero-momentum coordinate transformations is in fact much larger and an infinite number of adiabatic modes can be obtained in a similar way. However, except for the uniform and the gradient mode the rest involve primordial tensor modes \cite{Hinterbichler}.} 
Thus, one expects that unless there is an initial correlation between the long mode and short modes (leading to a violation of the Maldacena's consistency condition), a gradient mode should not result in any observable effect since it can be eliminated from the entire horizon by a single coordinate transformation \eqref{diff12}. In particular, the observed hemispherical power asymmetry \eqref{A} must be corrected to depend only on the deviation $f_{NL} - 5(1-n_s)/6$. 

Below, we will explicitly verify these expectation by deriving a second order expression for CMB multipoles, valid in the limit when at least one of the modes is super-horizon at recombination time. The formula can be useful in practice as it allows to derive the observable effect of the long mode on the short modes when the long mode enters the horizon after the recombination. 

\section{\label{SW}Sachs-Wolfe effect at second order}

We are ultimately interested in the effect of a long wavelength primordial fluctuation on the shorter wavelength CMB anisotropies. It helps to divide the problem into two phases: (a) Evolution of the primordial short scale perturbations in the background of the long mode until recombination, which we idealize as a hypersurface of constant average temperature $T_{\rm rec}$ at which the CMB photons are released, and freely propagate afterward. And (b) the effect of the long mode on the observed temperature and direction of the CMB photons. 

The first phase will be dealt with in the next section. The second phase is captured by the Sachs-Wolfe formula. It relates $T_o(\bsb{\hat n})$, the black body temperature of CMB photons observed in the direction $\bsb{\hat n}$, to the fluctuations of temperature, plasma velocity, and metric fluctuations at (and along the line of sight to) an earlier time-slice $\eta_e$--which though we refer to it as the emission time doesn't have to be so. Since we are interested in the long-short effects the formula must be derived to second order. This is done in appendix \ref{appSW} closely following \cite{Creminelli_squeezed} but keeping track of the observer's peculiar velocity $\bsb v_o$ and gravitational potentials $\Phi_o$ and $\Psi_o$. For simplicity the whole discussion is framed in a matter-dominated universe. The photon trajectories are solved for in appendix \ref{appSW} in the Poisson gauge \cite{Bertschinger}
\be
\label{metric}
ds^2 = a^2(\eta)[-e^{2\Phi}d\eta^2 +2\omega_i dx^i d\eta +(e^{-2\Psi}\delta_{ij}+\gamma_{ij})dx^i dx^j].
\ee
The spatial curvature and primordial tensor modes are neglected, hence $\omega_i$ and $\gamma_{ij}$ both start at second order in perturbations, and they are transverse ($\d_i\omega_i=0,\d_i\gamma_{ij}=0$). The result for CMB temperature in direction $\n$ is
\be
\label{theta}
\begin{split}
\frac{T(\bsb{\hat n})}{\bar T} -1
=&\Theta_e +\Phi_e+\bsb{\hat n}\cdot (\bsb v_o-\bsb v_e)+I-{\delta \n}\cdot \bsb v_e +
\bsb{\hat n} \cdot (\bsb v_o-\bsb v_e)(\Theta_e+\Phi_e)\\
&+\Phi_e\Theta_e +\frac{1}{2}\Phi_e^2
+(\bsb{\hat n} \cdot (\bsb v_o-\bsb v_e))^2-\frac{1}{2}(\bsb v_o-\bsb v_e)^2.
\end{split}
\ee
Let us define various quantities in this equation. The quantities with an index $e$, are evaluated at $(\eta_e,\x_e)$. In general $\eta_e$ does not have to be the same in all directions. As becomes clear in the following, for the emission hypersurface to have a uniform average temperature $T_{\rm rec}$, its geometry would be modified by long wavelength perturbations. Hence, $\eta_e$ will generically varies as a function of the local average of $\Phi_e$ over patches of size larger than the recombination horizon. Moreover, the (perturbed) position of the emission point is given to first order by
\be
\label{xr}
\bsb x_e = \bsb x_o + (\bsb{\hat n} -\bsb v_o + \bsb{\hat n} \bsb{\hat n}\cdot \bsb v_o)(\eta_o -\eta_e) + 2\bsb{\hat n} \int_{\eta_e}^{\eta_o} d\eta \Phi 
- 2\int_{\eta_e}^{\eta_o} d\eta (\eta-\eta_e) \nabla_\perp \Phi.
\ee
The quantities with index $o$ belong to the observer. The observer's position $\x_o$ is retained since it will be affected by the long gradient mode. Note that the time integrals are along the unperturbed photon trajectory $(\eta, \n(\eta_o-\eta))$. The three-velocity $\v$ is defined in terms of the four-velocity by $v^i\equiv a e^{-\Psi}u^i$. The Integrated Sachs-Wolfe (ISW) term,
\be
I(\eta_e,\eta_o)=\int_{\eta_e}^{\eta_o} d\eta 
\left(\Phi'+\Psi'+\omega_i'{n}^i-\frac{1}{2}\gamma'_{ij} { n}^i { n}^j\right),
\ee
is a second order effect in matter-dominance since at linear order $\Phi=\Psi = \rm{const.}$, and we are neglecting primordial tensor modes. At second order this is still irrelevant for the long-short effect \cite{Creminelli_SW}, but will be important in the calculation of long-long effect. The lensing $\delta\n$ is given by
\be\label{dn}
{\delta {\n}} = -\int _{\eta_e}^{\eta_o} d\eta \nabla_\perp (\Phi+\Psi),\qquad \text{with}\quad  \nabla_\perp = (\nabla- \n  \n\cdot\nabla).
\ee
Temperature contrast is defined as 
\be
\label{Tr}
\Theta (\eta_e,x_e)= \frac{T(\eta_e,x_e)}{\expect{T(\eta_e,\x)}_{\x}}-1,
\ee
where the average is over the whole time-slice $\eta_e$. Note that although $\eta_e$ generically fluctuates in the presence of long wavelength perturbations, this average is independent of them since any physical perturbation must vanish at infinity. Finally, $\bar T$ is defined by
\be\label{Tbar}
\bar T \equiv \frac{a_e}{a_o}e^{-\Phi_o} \expect{T(\eta_e,\x)}_{\x}.
\ee

Equation \eqref{theta} is the main result of this section. To study long-short effects one keeps the cross terms with one short- and one long-wavelength modes in the quadratic terms, including the effect of the long mode on the emission coordinate $\x_e$. In addition, the long mode modifies various averages and the geometry of the recombination surface which have to be taken into account. This will be discussed next.

\subsection{Observed average temperature and the recombination time}

\begin{figure}[t]
\centering
\includegraphics[scale = 0.8]{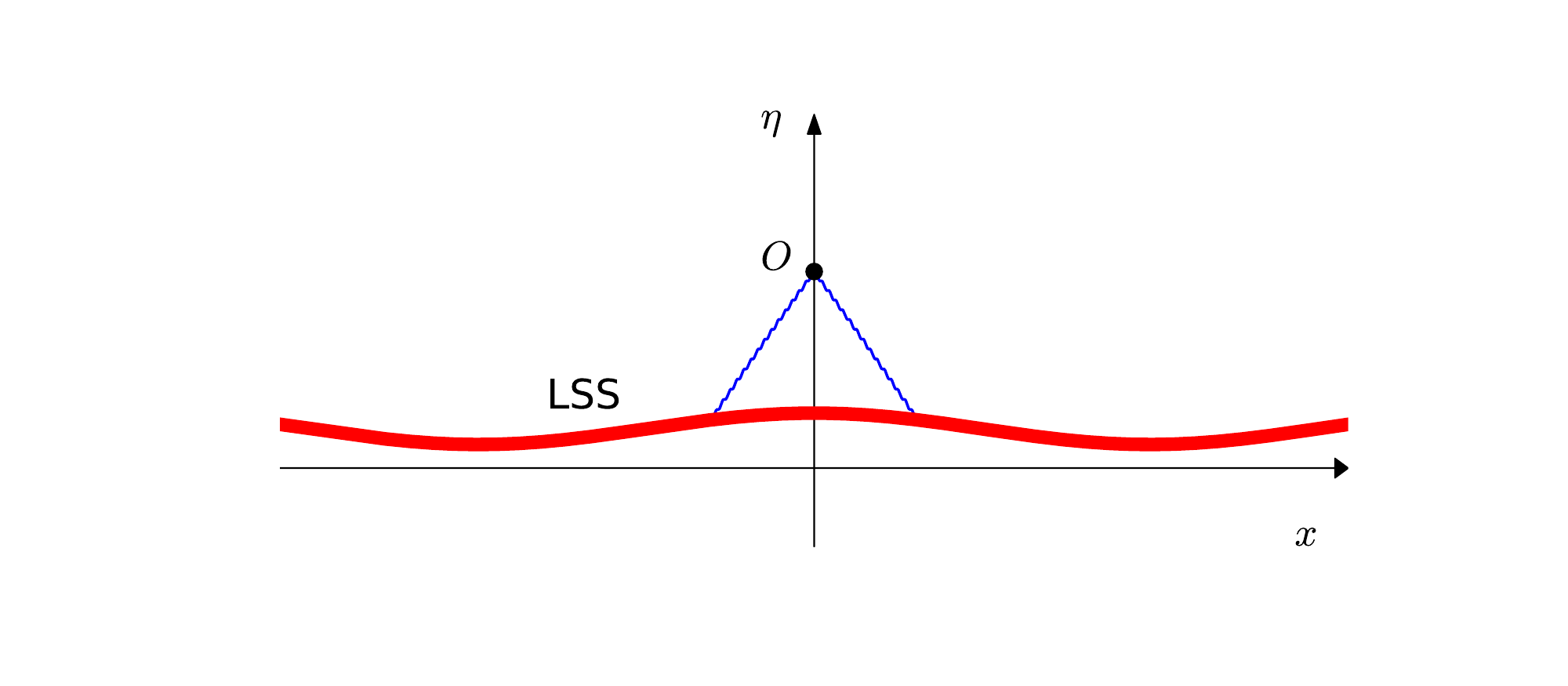} 
\caption{\small{A long-wavelength mode which does not average to zero in the observable patch of universe modifies the relation between $\expect{T_o(\n)}$ and $\bar T$, but also the global observation time $\eta_o$, according \eqref{etao}.}}\label{uniflss}
\end{figure}

Super long-wavelength modes, like the one in figure \ref{uniflss}, enter the relation between globally defined averages like $\expect{T(\eta_e,\x)}_{\x}$ and the locally measurable ones like $T_{\rm CMB}=\expect{T(\n)}$, which appear in the definition of temperature anisotropy 
\be
\label{Theta_def}
\Theta (\bsb{\hat n})\equiv \frac{T_o(\bsb{\hat n})}{\expect{T_o(\n)}}-1.
\ee
To express the l.h.s. of \eqref{theta} in terms of $\Theta(\n)$ one should relate $\expect{T(\n)}$ to $\bar T$. At linear level $\expect{T(\n)}$ is given by
\be\label{To}
\expect{T_o(\n)}= \frac{a_e}{a_o}e^{-\Phi_o} \expect{T(\eta_e,\x_e)}_{x_e} (1+{\bar \Phi_e}),
\ee
where $\bar\Phi_e \equiv\expect{\Phi(\eta_e,\x_e)}_{x_e}$. The average is over the emission points at the intersection of observer's past lightcone and the $\eta_e$ hypersurface. (This can differ from $\expect{\Phi(\eta_e,\x)}_\x$ in the presence of super-long fluctuations.) The last factor in \eqref{To} accounts for a uniform redshift. As is shown in \eqref{Tphi} below, the average temperature over the observable patch of sky $\expect{T(\eta_e,x_e)}_{x_e}$ in the presence of a gravitational potential that is uniform over the patch (or an average potential $\bar\Phi_e$) can be expressed in terms of the global average as $\expect{T(\eta_e,\x)}_{x_e}=\expect{T(\eta_e,\x)}_{\x}(1-2\bar\Phi_e/3)$. Substituting in \eqref{To} and using \eqref{Tbar} gives
\be
\expect{T_o(\n)}=\bar T (1+{\frac{1}{3}\bar{\Phi}_e}).
\ee
So the l.h.s. of \eqref{theta} is
\be
\label{thetaPhi}
\frac{T(\bsb{\hat n})}{\bar T} -1=\Theta(\n)(1+\frac{1}{3}\bar\Phi_e)+\frac{1}{3}\bar\Phi_e.
\ee

However, to use \eqref{theta} one should also take into account the slightly more subtle dependence of $\eta_o$ and $\eta_e$ on long-wavelength perturbations. The physically meaningful quantity to calculate is the expression of temperature anisotropy $\Theta(\n)$ at a fixed physical time $t_*$, measured by observer's clock, in terms of quantities at a surface of fixed average temperature $T_{\rm rec}$. In a matter dominated universe where the scale factor is given by $a=\eta^2$, the expression \eqref{metric} for the metric implies that in the presence of a uniform potential $\bar\Phi_e$ a constant proper time $t_*$ corresponds to
\be
\label{etao}
\eta_o=\eta_* (1-\frac{1}{3}\bar\Phi_e)
\ee
where $\eta_* = (3 t_*)^{1/3}$. Similarly, in the presence of a gravitational potential $\Phi_L$ whose wavelength exceeds the horizon at recombination time (but can be much shorter than the current horizon) the initial time slice $\eta_e$ should be perturbed (see figure \ref{osclss}). In a patch where $\Phi_L$ is approximately constant, we should take
\be
\label{etarec}
\eta_e = \eta_{\rm rec} (1-\frac{1}{3}\Phi_L)
\ee
if we are going to express the results in terms of quantities on a constant temperature slice. ($\eta_{\rm rec}$ is the constant time corresponding to constant temperature $T_{\rm rec}$ in a homogeneous cosmology.) These corrections would also modify the position of the emission point through \eqref{xr}.

\section{\label{LS}The long-short effects}

Having connected the CMB observables to perturbations on the recombination surface, we still need to understand how short scale fluctuations on this surface depend on the long wavelength mode. The dependence, which arises from the prior evolution of the short modes in the background of the long mode, introduces new short-long effects upon substitution in the linear terms of \eqref{theta}.

This problem greatly simplifies when the long mode is super-horizon at the recombination time since, as explained in section \ref{null}, the effect of a long adiabatic mode on the evolution of shorter modes is equivalent to the application of a coordinate-transformation \cite{Weinberg}. We use this to find an analytic second order expression for $\Theta(\n)$ in terms of linearly evolved short-wavelength quantities on the unperturbed recombination surface, and the long wavelength mode on and along the trajectory to that surface. We assume that the short modes are in vacuum when the long mode freezes out, and hence the Maldacena's consistency condition is satisfied. Any deviation should be added on top of our results.

To find the appropriate transformation substitute $a=\eta^2$ in \eqref{diff13} and \eqref{diff12}. This implies that a long mode $\Phi_L$ and its first derivative can be locally eliminated in the vicinity of a point $\bsb D$ on the recombination surface by the following coordinate transformation
\bea
\label{diff}
\tilde\eta = (1+\frac{1}{3}\Phi_L(\bsb x))\eta,~~~~~~~~~~~~~~~~~~~~~\\
\label{diffx}
\bsb{\tilde y}=(1-\frac{5}{3}\Phi_L(\bsb x))\bsb y+\frac{5}{6} y^2 \bsb q 
+\frac{1}{6}\eta^2 \bsb q,
\eea
where $\bsb y = \bsb x -\bsb D$, and
\bea
\label{phi}
\Phi_L(\bsb x)=\Phi_L(\bsb D)+\bsb q \cdot \bsb y +\O(y^2/\lambda_L^2),\qquad 
\bsb q = \nabla \Phi_L(\bsb D).
\eea
Note that in writing \eqref{diff} and \eqref{diffx} we are neglecting the $\O(y^2/\lambda_L^2)$ terms; the final result is, therefore, valid up to corrections of order $(\lambda_S/\lambda_L)^2$. Under this transformation the gravitational potential shifts non-linearly:
\be
\label{phi1}
\Phi(\eta,\bsb x)=\tilde\Phi(\tilde\eta,\bsb{\tilde {x}})+\Phi_L,
\ee
while temperature $T$ (and other scalar quantities) transform as
\be
T(\eta,\bsb x)=\tilde T(\tilde\eta,\bsb{\tilde {x}}).
\ee
The quantities with a tilde consists entirely of the short modes, and for our purposes can be assumed to be linearly evolved. To write temperature anisotropy $\Theta$ in the new coordinates we also need to find the new average temperature (see \eqref{Tr}). This can be obtained by noticing that the average temperature $\expect{T(\eta_e,\x)}_\x$ on the entire time-slice $\eta_e$ is independent of physical perturbations which die off at infinity. Hence, for fixed $\eta_1$ and $\eta_2$ we have 
\be
\expect{T(\eta_1,\bsb x)}=\expect{\tilde T(\eta_1,\bsb x)}=\expect{\tilde T(\eta_2,\bsb x)}\frac{a(\eta_2)}{a(\eta_1)},
\ee
as an exact relation. Note that while the first average must be over a large region of size $r\gg 1/q$ (so that the long mode averages to zero), the second average on $\tilde T$ can be over just a few patches of size $\cH_{\rm rec}$ since the tilde universe is empty from long-wavelength perturbations. In matter domination, $a(\eta_2)/a(\eta_1) = (\eta_2/\eta_1)^2$ and we get
\be
\label{Tphi}
\expect{T(\eta,\bsb x)}=\expect{\tilde T(\tilde \eta,\bsb x)} e^{2\Phi_L/3}.
\ee
Note that we have exponentiated \eqref{diff}, which is needed for the analysis of long-long effects in the next section. The justification in that case is that although the long gradient mode varies significantly across present horizon, it is almost constant over each horizon-size patch at recombination, and the exponentiation is valid for a constant $\Phi$. This gives
\be
\label{T2}
\Theta(\eta,\bsb x)=
(\tilde\Theta(\tilde\eta,\tilde \x)+1)e^{-2\Phi_L/3}-1.
\ee
Finally, the plasma velocity transforms as follows (apart from the usual change of the arguments $(\eta,\x)\to (\tilde \eta,\tilde \x)$)
\be
u^i = \frac{\d x^i}{\d {\tilde x}^\mu}{\tilde u}^\mu = (1+\frac{5}{3}\Phi_L){\tilde u}^i+\frac{5}{3}(y^i \bsb q\cdot \bsb u -q^i \bsb q \cdot \bsb y)-\frac{1}{3a} q^i \eta_e.
\ee
Thus in a matter dominated universe
\be
\label{v}
\bsb v=ae^{-\Psi}\bsb u=\bsb{\tilde v} +\frac{5}{3}(\bsb{\tilde v}\cdot \bsb q \; \bsb{y} - \bsb{\tilde v}\cdot \bsb{y}\; \bsb q)-\frac{1}{3}\eta_e \bsb q.
\ee
Substituting equations (\ref{phi1},\ref{T2},\ref{v}) in \eqref{theta} and \eqref{thetaPhi}, and setting $\bar\Phi_e=0$ for the moment, we get
\be
\label{theta1}
\Theta(\bsb{\hat n})=\Theta_{S,\rm obs}+\Theta_{L,\rm obs}+\Theta_{L,\rm obs}\Theta_{S,\rm obs}+\Delta\bsb x_e\cdot \nabla \Theta_{S,\rm obs} - \Delta\n \cdot \tilde{\bsb v}
\ee
where
\bea
\label{TS}
\Theta_{S,\rm obs}=\tilde \Theta+\tilde \Phi-\bsb{\hat n}\cdot \bsb{\tilde v}~~~~~~~~~~~\\
\label{TL}
\Theta_{L,\rm obs}=\frac{1}{3}\Phi_L+\frac{1}{3}\eta_e \bsb{\hat n}\cdot \bsb q +\bsb{\hat n}\cdot \bsb v_o,
\eea
are respectively the observed short- and long-wavelength $\Theta(\n)$ at linear order. The aberration is given by 
\be
\label{Dn}
\Delta\n =\delta \n+(\bsb{\hat n}\cdot \bsb v_o+\frac{1}{3}\eta_e \bsb{\hat n}\cdot \bsb q  - \frac{5}{3} \bsb y \cdot \bsb q)\bsb{\hat n} 
-(\bsb v_o+\frac{1}{3}\eta_e \bsb q  - \frac{5}{3}\n\cdot \bsb y \;\bsb q),
\ee
with $\delta\n$ given in \eqref{dn}. Moreover, since the short-wavelength quantities are evaluated at $\bsb{\tilde x}_e$ as given by \eqref{xr} and \eqref{diffx}, to get the full second order expression one should also expand around the unperturbed recombination position which introduces an additional second order term with 
\be
\label{delT}
\begin{split}
\Delta \bsb x_e=&\frac{1}{3}\Phi_L \eta_e \bsb{\hat n}+\Delta \eta(-\bsb v_o  +\bsb{\hat n} \bsb{\hat n}\cdot \bsb v_o)+2\bsb{\hat n} \int_{\eta_e}^{\eta_o} d\eta \Phi - 2\int_{\eta_e}^{\eta_o} d\eta (\eta-\eta_e) \nabla_\perp \Phi \\
&-\frac{5}{3}\Phi_L \bsb x +\frac{5}{6} y^2 \bsb q +\frac{1}{6}\bsb q \eta_e^2
\end{split}
\ee
where $\Delta \eta$ is the zeroth order $\eta_o - \eta_e$, and the first term on the r.h.s. is responsible for the shift 
\be
\n(\eta_o-\eta_e)\to \n(\eta_o-\tilde \eta_e)
\ee
in the argument of short quantities: $\Theta_{S,\rm obs}(\tilde \eta_e,\bsb x_o +\bsb{\hat n}(\eta_o-\tilde \eta_e))$. That is, we are expressing the observed $\Theta (\bsb{\hat n})$ in terms of quantities at the surface of constant temperature $T_{\rm rec}$ and not a surface of constant $\eta_e$. In the presence of the long mode $\Phi_L$ the recombination time $\eta_e$ itself fluctuates and as showed in \eqref{etarec} the constant temperature slice corresponds to $\tilde \eta_e=\eta_{\rm rec}=\rm const$. Hence, there is no time-derivative term in \eqref{theta1} in contrast to the result of \cite{Creminelli_squeezed}; otherwise, the two expressions agree when $\nabla\Phi_L$ is neglected.

The coordinate transformation \eqref{diff} takes care of the effect of the long mode on the short modes at recombination. The long wavelength can still be much shorter than the present horizon, in which case, the contribution to observer's position $\x_o$ and velocity $\v_o$, as well as the integrated effects in $\Delta\n$ and $\Delta\x_e$ along the photon trajectory would depend on the dynamical evolution. Equation \eqref{theta1} is the main result of this section. It expresses CMB multipoles in terms of the linearly evolved fields up to corrections of order $(\lambda_S/\lambda_L)^2$.\footnote{The contributions to $\x_o$ and $\v_o$, though important for the consistency check, are practically degenerate with the overall normalization and the observer's peculiar velocity.}

In the extreme case when the long wavelength is much longer than the present horizon, $\Phi_L$ is just a constant plus a pure gradient and the transformation \eqref{diff} removes it from the entire observable universe. Hence, we can perform a consistency check of the above formalism by verifying that such a mode makes no contribution to $\Theta(\n)$ measured at a fixed physical time $t_*$.

\subsection{Pure gradient consistency check}

\begin{figure}[t]
\centering
\includegraphics[scale = 0.8]{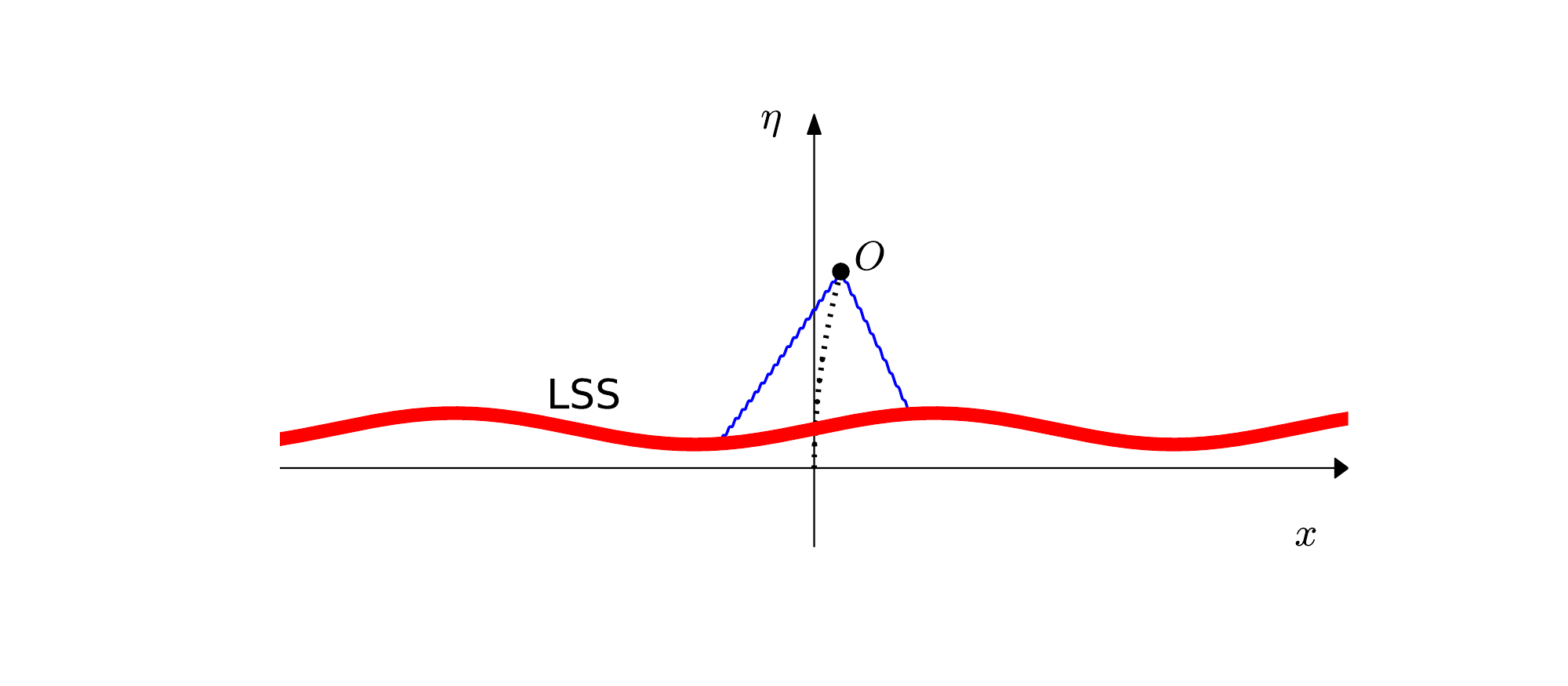} 
\caption{\small{A pure gradient mode accelerates the observer and tilts the recombination surface.}}\label{gradlss}
\end{figure}

In practice, it is easier to center the coordinate transformations \eqref{diff} and \eqref{diffx}, used to derive the long-short effects, close to the observer's position $\bsb x =0$ rather than a point on recombination surface ($\x =\bsb D$). They become
\bea
\label{diff1}
\tilde\eta = (1+\frac{1}{3}\Phi_L(\bsb x))\eta,~~~~~~~~~~~~~~~~~~~~~\\
\label{diff1x}
\bsb{\tilde x}=(1-\frac{5}{3}\Phi_L(\bsb x))\bsb x+\frac{5}{6}x^2 \bsb q 
+\frac{1}{6}\eta^2 \bsb q,
\eea
where $\Phi_L(\eta,\bsb x)$ is time-independent to first order, and is given by
\be
\label{phi2}
\Phi_L = \bar\Phi_e + \bsb q\cdot \bsb x.
\ee
This changes $\bsb y \to \bsb x=\n \Delta\eta +\O(\Phi_L)$ everywhere in \eqref{Dn} and \eqref{delT}. Equation \eqref{v} now changes to
\be
\bsb v=\bsb{\tilde v} +\frac{5}{3}(\bsb{\tilde v}\cdot \bsb q \; \bsb{x} - \bsb{\tilde v}\cdot \bsb{x}\; \bsb q)-\frac{1}{3}\eta_e \bsb q.
\ee
Now we can use this formula with $\eta_e\to \eta_o$ and $\x \to 0$ to derive the effect of the gradient mode on the observer's velocity and position (figure \ref{gradlss}):
\be
\label{vo}
\bsb v_o=\tilde{\bsb v}_o-\frac{1}{3}\eta_o \bsb q.
\ee
and $\bsb x_o$ changes according to \eqref{diff1x}. We assume $\tilde {\bsb x}_o=0$ and $\tilde {\bsb v} _o=0$, so
\be
\label{xo}
\bsb x_o = -\frac{1}{6}\eta_o^2\bsb q. 
\ee

Furthermore, the constant potential $\bar\Phi_e$ across the entire horizon implies that the l.h.s. of \eqref{theta1} must be replaced with \eqref{thetaPhi}, and the conformal time of observation $\eta_o$ be appropriately rescaled as in \eqref{etao}. This modifies $\Delta\x_e$ since now the unperturbed position on the recombination surface is $\n (\tilde\eta_o-\tilde\eta_e)$. Including this and the change \eqref{xo} in \eqref{delT}
\be
\Delta \bsb x_e \to \Delta \bsb x_e -\frac{1}{3}\bar \phi_e \eta_o \n -\frac{1}{6}\eta_o^2\bsb q
\ee
and substituting (\ref{phi2},\ref{vo}) we get $\Delta \bsb x_e =0$. The lensing $\delta \n$ in \eqref{Dn} evaluates to
\be
\delta \n = -2\Delta \eta (\bsb q - \bsb q\cdot \n \; \n),
\ee
resulting in $\Delta \n =0 $. Finally, equation \eqref{TL} simplifies to $\Theta_{L,\rm obs} = \bar \Phi_e/3$ and leads to 
\be
\Theta(\n) = \Theta_{S,\rm obs},
\ee
as desired. 

The cancellation of the effect of a constant mode has been shown in \cite{Creminelli_squeezed}, and the cancellation of the contribution of a pure gradient to the dipole moment in \cite{Erickcek_dipole}. But the absence of any modulation of the short modes by the linear gradient is new -- it would have resulted in a dipolar power asymmetry.

If the Maldacena's consistency condition is violated, it corresponds to a correlation between the initial conditions of the short and the long modes, and hence a physical local non-Gaussianity $\tilde f_{NL}=f_{NL}-5(1-n_s)/6$. Therefore, after removing the dynamical effect of the long mode by a coordinate transformation one obtains
\be
\tilde \Phi = g_S (1+ \tilde f_{NL}\: g_L),\qquad \text{with}\qquad \tilde f_{NL}=f_{NL}-5(1-n_s)/6,
\ee
which results in
\be
\Theta(\n) = \Theta_{g_{S},\rm obs}(1+\tilde f_{NL} \bsb q \cdot \bsb x),
\ee
This should replace equation \eqref{A} for the hemispherical power asymmetry in the presence of a linear gradient mode.
\section{\label{long}The long-long effects}

So far we saw that a linear gradient mode has no observable effect, except for a possible modification of the initial condition for the short modes. Having derived the second order Sachs-Wolfe formula, it is natural to ask whether this continues to be the case to second order in the long mode. Now one needs to keep terms quadratic in the long mode and must take into account the second order time-evolution of a linear gradient initial condition. Clearly, this no longer reduces to performing a coordinate transformation on the linearly evolved solution, and we will use the existing second order solution of \cite{Matarrese,Bartolo,Creminelli_action}. In the following, we will first find the second order effect of a gradient initial condition in section \ref{MQ}. There is a finite contribution to the monopole and quadrupole moments. We will see that these contributions would be cancelled if the linear gradient was superposed with a parabolic primordial perturbation, thereby leaving no observable effect. We will give an interpretation of this special superposition in terms of a non-linear extension of Weinberg's adiabatic modes in section \ref{2nd}.

\subsection{Monopole and quadrupole\label{MQ}}

A linear gradient mode that vanishes at our location $\x=0$, should be thought of as the small $\x$ limit of a super-horizon physical mode, e.g.
\be
\Phi= \phi_k \sin \k\cdot \x,\qquad \q = \phi_k \k,\qquad kD\ll 1,
\ee
where $D$ is the size of the observable universe. For any such mode if the amplitude $\phi_k$ satisfies $kD\ll\phi_k\ll 1$, the physical effects arising from the linear curvature of the mode which are of order $D^2 \nabla^2 \Phi \sim\phi_k k^3 D^3$ are negligible compared to non-linear effects $D^2 (\nabla\Phi)^2\sim \phi_k^2 k^2 D^2$. Therefore, it is meaningful to talk about a gradient mode at second order. At this order there are monopole and quadrupole contributions. For simplicity we also neglect $\eta_e/\eta_o$. This implies that the modes we are interested in are so far outside the recombination horizon that we can neglect $\bsb v$ in the Sachs-Wolfe formula. Moreover, although we are interested in the variation of the gradient mode over the present horizon, in the same approximation we can treat it effectively as a piecewise uniform mode on several super-horizon patches of the recombination surface. This allows us to exponentiate the coordinate transformation \eqref{diff} in each patch:
\be
\eta_e = \eta_{\rm rec} e^{-\Phi_e/3}.
\ee
Here, we chose a constant $\tilde\eta=\eta_{\rm rec}$. Since there is no short-wavelength perturbation, with this choice the temperature on the emission hypersurface would be uniform $T(\eta_e,\x_e) = T_{\rm rec}$. Therefore, using this and plugging $a_e=\eta_e^2$ and \eqref{tilden} in \eqref{To/Te}, and expanding to second order, we obtain
\be
\label{T(n)}
T(\bsb{\hat n})
=\frac{a_{\rm rec}}{a_o} e^{-\Phi_o}T_{\rm rec}\left[1+\bsb{\hat n}\cdot \bsb v_o
+\frac{1}{3}\Phi_e(1+\bsb{\hat n}\cdot \bsb v_o)+\frac{1}{18}\Phi_e^2
+I+(\bsb{\hat n} \cdot \bsb v_o)^2-\frac{1}{2}(\bsb v_o)^2\right]\\[10pt]
\ee
where $\Phi_e=\Phi(\eta_e,\x_e)$ is the second order solution resulting from a linear gradient initial condition and as before the emission point $\x_e$ is given by \eqref{xr}. (This formula agrees with \eqref{theta} after substituting \eqref{T2} for $\Theta_e$ and setting $\tilde\Theta=0$, which corresponds to the absence of short modes.) 

To calculate $T(\n)$ we need the knowledge of the second order solution for metric perturbations. In a matter dominated universe this has been calculated in \cite{Matarrese,Bartolo,Creminelli_action} (we use the results of the latter). The result is of the form
\be
\begin{split}
\label{PhiPsi2}
\Phi =& \phi +A+B\eta^2\\[10pt]
\Psi =& \phi -\frac{2}{3}A+B\eta^2,
\end{split}
\ee
where $\phi$ is the initial condition, proportional to the comoving-gauge scalar curvature ($\phi=-3\zeta/5$). In general, it may contain a constant second order piece: $\phi = \phi_1+\phi_2$. We are eventually interested in $\phi_1=\q\cdot\x$, but will have more to say about $\phi_2$. $A$ and $B$ are given by the non-local expressions:
\bea
A & =  &\left[\d^{-2}(\d_j\phi)^2 -3\d^{-4}\d_i\d_j(\d_i\phi\d_j\phi) \right]\\[10pt]
\label{B}
B  & = &\frac{1}{12}(\d_j\phi)^2+\frac{5}{42}\d^{-2}\left[(\d^2\phi)^2-(\d_i\d_j\phi)^2\right] \,.
\eea
The expressions for $\omega_i$ and $\gamma_{ij}$ are given in appendix \ref{second}. They are also non-local, and are respectively $\O(\eta)$ and $\O(\eta^2)$ in the long-wavelength limit since the primordial tensor component is neglected. 

The non-locality of the solution implies that these quantities depend on how exactly the gradient mode has been extended to a physical mode beyond our horizon. Observable quantities such as CMB multipoles cannot depend on this, so the non-localities in these expressions must cancel in \eqref{T(n)}. This is shown in appendix \ref{local}. Neglecting $\eta_e/\eta_o$, the non-local terms contribute to linear SW, ISW term, and also to observer's potential $\Phi_o$ and velocity $\v_o$. The former two combine to give
\be
\label{Phi+I}
\frac{1}{3}\Phi_e+I= \frac{1}{3}\phi(\x_e) 
-\frac{1}{3}\int d\eta \eta [2(\bsb{\hat n} \cdot \nabla \phi)^2-(\d_j \phi)^2]+\Theta_0+\Theta_1 + \cdots
\ee
where dots correspond to local, higher derivative terms that vanish on $\phi_1 = \q\cdot \x$. (Recall that $\phi$ is the initial condition and time-independent.) $\Theta_0$ and $\Theta_1$ are non-local expressions which would naively contribute to the monopole and dipole moments. However, as already seen in the analysis of the long-short effect the naive expressions for monopole and dipole are degenerate with the observer's time $\eta_o$ and velocity $\v_o$. In appendix \ref{mondi} we show that the above ambiguous results for $\Theta_0$ and $\Theta_1$ are locally equivalent to a coordinate transformation and they cancel with a similar contribution of the non-local pieces in $\Phi$ to observer's time and velocity. 

Since all non-localities cancel we can unambiguously calculate the second order contribution of a gradient mode to $T(\n)$. Substituting $\phi(\x_e) = \q\cdot\x_e +\phi_2$ in \eqref{Phi+I}, with $\x_e$ calculated to first order from \eqref{xr}, and discarding $\Theta_{0,1}$ yields
\be
\begin{split}
&T(\bsb{\hat n})
=\frac{a_{\rm rec}}{a_o} e^{-\Phi_o}T_{\rm rec}\Big[1-\frac{1}{6}q^2\eta_o^2
+\frac{5}{18}(\bsb q\cdot \bsb x)^2+\frac{1}{3}\phi_2\\[10pt]
&~~~~~~+(\frac{1}{3}\bsb q\cdot \bsb x+\bsb{\hat n}\cdot \bsb v_o )+(\frac{1}{3}\bsb q\cdot \bsb x+\bsb{\hat n}\cdot \bsb v_o )^2-\frac{1}{2}(\frac{1}{3}\bsb q\eta_o+\bsb v_o)^2
+\frac{1}{3}(\frac{1}{6}q^2\eta_o^2+\q \cdot \x_o)^2\Big],
\end{split}
\ee
where $\x = \eta_o \n$. The contribution of the long mode to the second line vanishes since
\be
\label{vo1}
\v_o = -\frac{1}{3}\q\eta_o,\qquad \x_o= -\frac{1}{6}\q \eta_o^2.
\ee
(We assumed the unperturbed observer is at rest at $\tilde\x_o =0$.) However, there is an extra contribution to monopole from the pre-factor $ e^{-\Phi_o}/a_o$. First, notice that according to \eqref{PhiPsi2} there is a time-dependent piece $q^2\eta^2/12$ in $\Phi$ coming from the local term in \eqref{B}. We therefore have
\be
\label{Phio}
\Phi_o = \tilde\Phi_o + \q\cdot\x_o +\frac{1}{12}q^2\eta_o^2 = \tilde\Phi_o - \frac{1}{12}q^2\eta_o^2.
\ee
To keep fixed the observer's proper time $t$, the time-like geodesic equation in matter-dominated universe should be used to derive the modified $\eta_o$:
\be
\frac{dt}{d\eta_o} = \eta_o^2 e^{\Phi_o}\sqrt{1-v_o^2}.
\ee
Using \eqref{vo1} and \eqref{Phio}, and ignoring the short wavelength $\tilde\Phi$, this can be integrated to give
\be
t = \frac{1}{3}\eta_o^3\left(1-\frac{1}{12}q^2\eta_o^2\right).
\ee
For a constant $t=t_*$, the conformal time of observation is rescaled by $\eta_o = \eta_*(1+q^2\eta_o^2/36)$. Combining all pieces we finally obtain
\be
T(\bsb{\hat n})
=T_{\rm CMB}\Big[1-\frac{5}{36}q^2\eta_o^2+\frac{5}{18}(\bsb q\cdot \bsb x)^2+\frac{1}{3}\phi_2\Big].
\ee
A pure gradient initial condition with $\phi_2=0$, therefore, leads to a finite contribution to the CMB monopole and quadrupole. In the presence of primordial local non-Gaussianity, $\tilde f_{NL}\neq  0$, one naturally expects a gradient mode to be accompanied by a parabolic piece which induces a quadrupole moment. The above result shows how the subsequent non-linear evolution and the transformation into CMB observable generates an additional contribution. This can be used to observationally constrain the amplitude of a pure gradient initial condition.

What can we say about the non-Gaussian piece $\phi_2$? In the context of single-field slow-roll inflation it is natural to take the primordial $\zeta_0$ to be an exact gradient, because its self-interactions are slow-roll suppressed and hence the Fourier modes are good approximate solutions even at non-linear level \cite{Maldacena}. This can then be translated to fluctuations in the Poisson gauge \eqref{metric}. For a uniform super-horizon perturbation (which is a valid assumption at the initial time $\eta =0$) we get
\be
\phi = -\frac{3}{5}\zeta_0
\ee
to all orders. Therefore, taking $\zeta_0$ to be a sinusoidal wave generated during inflation implies $\phi_2\simeq 0$. Hence the pure gradient mode, although unobservable at first order, induces non-zero monopole and quadrupole moments at second order. This constrains the amplitude of such a mode even in the absence of primordial non-Gaussianity $\tilde f_{NL}$.

However, if we superpose the linear gradient with $\phi_2 = 5q^2\eta_o^2/12 -5 (\bsb q\cdot \bsb x)^2/6$ the second order contribution of the long mode to the CMB temperature vanishes as well. We will next show that this combination corresponds to an adiabatic mode at second order: a physical mode that can be locally removed by a coordinate transformation.
 
\subsection{Gradient adiabatic mode at second order\label{2nd}}

It is natural to expect that the adiabatic modes of Weinberg exist beyond linear level, corresponding to slowly varying finite physical modes that can locally be approximated by a large (non-vanishing at infinity) coordinate transformation. As such, they must be unobservable to all orders. This is obviously the case for a uniform adiabatic mode: In the comoving gauge \eqref{h}, a finite constant $\zeta$ corresponds to a constant rescaling $x^i\to x^i e^{\zeta}$.\footnote{The result also exponentiates in the Poisson gauge with $\Phi=\Psi = -5\zeta/3$ up to an ambiguity corresponding to a time-redefinition.} In this subsection we construct the gradient adiabatic mode at second order, by demanding that the second order solution of previous subsection to be equivalent to a coordinate transformation. Although this non-linear extension is non-unique, due to the non-locality of the solution \eqref{PhiPsi2}, the second order initial condition $\phi_2$ is uniquely determined to make the adiabatic mode unobservable.

Following Weinberg \cite{Weinberg} we apply to an unperturbed FRW background the most general second order extension of the diffeomorphism ({\ref{diff1},\ref{diff1x}) with $\bar\Phi_e=0$
\be\label{diff2}
\begin{split}
\eta \;\to\; & (1-\frac{1}{3}\bsb q\cdot \bsb x)\eta +a_1 (\bsb q\cdot \bsb x)^2\eta+a_2q^2 x^2 \eta +a_3 q^2 \eta^3,\\[10pt]
\bsb x\;\to\; &(1+\frac{5}{3}\bsb q\cdot \bsb x)\bsb x -\frac{5}{6}\bsb q x^2 -\frac{1}{6}\bsb q \eta^2 + b_1(\bsb q\cdot \bsb x)^2\bsb x+b_2q^2 \eta^2 \bsb x +b_3 q^2 x^2 \bsb x \\[10pt]
& +b_4 x^2 (\bsb q\cdot \bsb x) \bsb q+b_5 \eta^2 (\bsb q\cdot \bsb x)  \bsb q.
\end{split}
\ee
We calculate the resulting metric perturbations and impose two requirements: (i) the Poisson gauge condition is preserved, and (ii) they matches the second order solution.

In matter-domination, we get the following expressions for various metric components:
\be
\label{Phi_diff}
e^{2\Phi}=1+2\bsb q\cdot \bsb x+(\frac{5}{3}+6a_1)(\bsb q\cdot \bsb x)^2+6a_2 q^2 x^2 +(7a_3-\frac{1}{9})q^2\eta^2,~~~~~~~~~~~~~~~~~~~
\ee

\be
\label{Psi_diff}
\begin{split}
e^{-2\Psi}=&1-2\bsb q\cdot \bsb x+(-1+4a_1+2b_1)(\bsb q\cdot \bsb x)^2+(4a_2+2b_3)q^2x^2+(4a_3+2b_2)q^2\eta^2~~~~~\\[10pt]
&+\frac{1}{3}q^2[(\frac{50}{9}+2b_4 +4b_3)x^2 +(-\frac{1}{9}+2b_5)\eta^2]~~~
+\frac{1}{3}(\bsb q\cdot \bsb x)^2(-\frac{50}{9}+4b_1+4b_4)
\end{split}
\ee

\be\label{omega_diff}
\omega_i = (-\frac{11}{9}-2a_1+2b_5)(\bsb q\cdot \bsb x) q^i \eta+(\frac{5}{9}-2a_2+2b_2)q^2\eta x^i,~~~~~~~~~~~~~~~~~~~~~~~~~~~~~~
\ee

\be\label{gamma_diff}
\begin{split}
\gamma_{ij}\;=\;& (q^i q^j -\frac{1}{3}q^2 \delta^{ij})[(\frac{25}{9}+2b_4)x^2 +(-\frac{1}{9}+2b_5)\eta^2]
+(\frac{25}{9}+4b_3) q^2 (x^i x^j -\frac{1}{3} x^2 \delta^{ij})q^2 \\[10pt]
&+[\frac{1}{2}(q^ix^j+x^iq^j)-\frac{1}{3}\bsb q\cdot \bsb x \delta^{ij}](\bsb q\cdot \bsb x)(-\frac{50}{9}+4b_1+4b_4).\\[10pt]
\end{split}
\ee
The coefficients $a_i$ and $b_i$ are not all independent. Requiring the new metric to satisfy the gauge condition $\d_i\omega^i=0=\d_i\gamma_{ij}$ puts four constraints on them. The matching with the physical solution seems impossible at first sight because of the non-locality of the latter. However, as already seen, those non-localities cancel in the observable CMB multipoles. So we fix only a subset of parameters based on the structure of the physical solution; the residual freedom corresponds to the redundancy of our description. In particular, in order for $\gamma_{ij}(\eta=0)=0$, one must set
\be
\label{bs}
b_1 = \frac{25}{9},\qquad b_3 = -\frac{25}{36}\qquad b_4=-\frac{25}{18}.
\ee
This is sufficient to determine the second order piece $\phi_2$. Comparing \eqref{Phi_diff} and \eqref{Psi_diff} with the expression \eqref{PhiPsi2} for $\Phi$ and $\Psi$ (with $\phi_1 =\q\cdot\x$), and using the above values for $b_{1,3,4}$ gives
\be
\begin{split}
A=&\left(3 a_2-\frac{5}{12}\right) q^2 x^2+(\frac{2}{3}+3a_1)(\bsb q\cdot \bsb x)^2\\[10pt]
\phi_2 =&\frac{5}{12}q^2x^2 -\frac{5}{6} (\bsb q\cdot \bsb x)^2 .
\end{split}
\ee
The ambiguity in $A$ corresponds to the fact that depending on how the gradient mode is extended to a physical mode at large distances the $\O(x^2)$ piece of the potentials $\Phi,\Psi$ vary. However, those ambiguities cancel as shown in appendix \ref{local}. On the other hand, the second order field $\phi_2$ is uniquely determined to be what is needed to get zero monopole and quadrupole moments.\footnote{In fact \eqref{diff2} is not the most general diffeomorphism. One should also include new $\O(x)$ and $\O(x^2)$ corrections, and hence new uniform and gradient components in metric perturbations that are respectively of order $\phi_k^2$ and $\phi_k^2 k x$. They obviously depend on how the gradient mode is extended at infinity since they depend separately on $\phi_k$ and $k$, and not just through $\q=\phi_k \k$. These ambiguous corrections naively lead to monopole and dipole contributions to $\Theta(\n)$, but as discussed in appendix \ref{mondi}, they cancel in the physical observables.}

We should finally mention that there are linearized adiabatic modes that start at linear order from $\mal O(x^2)$ or higher \cite{Hinterbichler}. However, all of them involve primordial tensor modes which were forbidden by setting $\gamma_{ij}=0$ at initial time and imposing \eqref{bs}. Projecting cosmological correlation functions on the adiabatic modes leads to consistency conditions similar to the one reviewed in section \ref{null} (see, e.g., \cite{Creminelli_sct,Hinterbichler_adia,Hinterbichler}). There are double-soft consistency conditions which arise from our the second order extension of adiabatic modes. They are dicussed in \cite{double_soft,Joyce}.

\section{Conclusions}

In this paper we argued that a linear gradient mode does not induce any dynamical effect because it is locally equivalent to a coordinate transformation \cite{Weinberg,Creminelli_sct}. In particular, this implies that when Maldacena's consistency conditions are satisfied, a long gradient mode extending across our horizon cannot lead to any hemispherical asymmetry in the CMB power spectrum (as suggested by, e.g., \cite{Erickcek}). Next, we derived a second order formula for the CMB multipoles in the squeezed limit in terms of the linearly evolved fields. The formula, which generalizes that of \cite{Creminelli_squeezed}, is valid as long as $\lambda_L$ is longer than the recombination horizon and up to corrections of order $(\lambda_S/\lambda_L)^2$, where $\lambda_{S,L}$ are the short and long wavelengths. The derivation rests crucially upon our symmetry arguments which reduces the effect of the long mode on the short modes until recombination time to a coordinate transformation. This analytic result can provide another check of the second order numerical Boltzmann codes besides the one performed in \cite{Creminelli_squeezed,Huang}. Moreover, it supports or general argument for the absence of any observable effect, such as power asymmetry, caused by a linear gradient mode whose wavelength is much longer than the present horizon size. Finally, we studied the second order effect of a gradient mode and showed that it induces observable monopole and quadrupole moments. We showed how it must be superposed by a quadratic primordial perturbation to become locally equivalent to a coordinate transformation up to second order. This provides an explicit example of a second order adiabatic mode, and leads to double soft consistency conditions \cite{double_soft,Joyce}.

\section*{Acknowledgments}

We thank Ali Akbar Abolhasani, Paolo Creminelli, and Mohammad Hossein Namjoo for stimulating discussions and comments. M.M. is supported by NSF Grants PHY-1314311 and PHY-0855425. M.Z. is supported in part by the NSF grants AST-0907969, PHY-1213563 and AST-1409709.
\appendix

\section{Second order Sachs-Wolfe formula\label{appSW}}

To derive the Sachs-Wolfe formula in our approximation we use the fact that for black-body radiation the observed temperature $T(\n)$ is rescaled with respect to the initial temperature $T(\eta_e,\x_e)$ by the amount of redshift photons experience:
\be
T(\bsb{\hat n}) = \frac{E_o}{E_e}T(\eta_e,\bsb x_e).
\ee 
The indices $o$ and $e$ indicate the observation and the emission events, respectively. The energy of the photon $E$ is given in terms of its null four-momentum $P^\mu=dx^\mu /d\lambda$, and observer's (emitter's) four-velocity $u^\mu$, by
\be
E=-P_\mu u^\mu.
\ee
In our convention $u^\mu u_\mu =-1$. In the Poisson gauge \eqref{metric}, the redshift becomes
\bea
\label{redshift}
\frac{E_o}{E_e}=\frac{a_e P_0(\eta_o)}{a_o P_0(\eta_e)}e^{\Phi_e - \Phi_o}
\left(\sqrt{1+v^2}+e^{\Phi+\Psi}\frac{P_i}{P_0}v^i\right)_o \left(\sqrt{1+v^2}+e^{\Phi+\Psi}\frac{P_i}{P_0}v^i\right)_e^{-1},
\eea
where $v^i\equiv a e^{-\Psi}u^i$, and $u^0=\sqrt{1+v^2}e^{-\Phi}/a$. Moreover, $\Phi_e \equiv \Phi(\eta_e,\x_e)$ and similarly for other fields. For a given direction $\bsb{\hat n}$ of the incident photon, $P_0$ and $P_i$ are determined by the geodesic equation
\be
\label{geo}
\frac{dP_\mu}{d\lambda}=\frac{1}{2}\d_\mu g_{\alpha\beta}P^\alpha P^\beta.
\ee
To second order, the zeroth component of this equation yields
\be
\label{dP0/dl}
\frac{1}{P_0}\frac{dP_0}{d\eta}=\Phi'+\Psi'+\omega_i'\bsb{\hat{\tilde n}}^i-\frac{1}{2}\gamma'_{ij}\bsb{\hat{\tilde n}}^i\bsb{\hat{\tilde n}}^j,
\ee
where $\bsb{\bsb{\hat{\tilde n}}}$ is related to the observation angle $\bsb{\hat n}$ and observers peculiar velocity $\bsb v_o$ via
\be
\label{tilden}
\bsb{\bsb{\hat{\tilde n}}}=\bsb{\hat n} -\bsb v_o + \bsb{\hat n} \bsb{\hat n}\cdot \bsb v_o.
\ee
We are only interested in the contribution of long-gradient mode to $\bsb v_o$ and regard it as a long-wavelength perturbation, hence at zeroth order $\bsb{\hat{\tilde n}}=\bsb{\hat n}$. In deriving \eqref{dP0/dl} we have used the first order null condition ${P^i}^2 = e^{2(\Phi+\Psi)}{P^0}^2$, and
\be
{{\tilde n}}^i = -\frac{P^i(\eta_o)}{P^0(\eta_o)}e^{-(\Phi_o+\Psi_o)}.
\ee
To zeroth order, $P^i/P^0=-{\tilde n}^i$ for all $\eta$. Integrating \eqref{dP0/dl} leads to the integrated Sachs-Wolfe effect:
\be
\label{I}
\frac{P_0(\eta_o)}{P_0(\eta_e)}=\exp[I(\eta_e,\eta_o)],\quad 
I(\eta_e,\eta_o)=\int_{\eta_e}^{\eta_o} d\eta 
\left(\Phi'+\Psi'+\omega_i'{\tilde n}^i-\frac{1}{2}\gamma'_{ij} {\tilde n}^i {\tilde n}^j\right).
\ee
The integral is along the photon trajectory $(\eta,\bsb{\hat{\tilde n}}(\eta_o -\eta))$. In the following we take $P_0(\eta_o)=1$, which to zeroth order remains conserved along the photon trajectory.

We will only need the first order solution for $P_i$, which from \eqref{geo} satisfies
\be
\frac{1}{P_0}\frac{dP_i}{d\eta}=\d_i(\Phi+\Psi).
\ee
Integrating this using the zeroth order value $P_0=1$ gives
\be
P_i(\eta_e)=P_i(\eta_o)-\int_{\eta_e}^{\eta_o} d\eta \d_i(\Phi+\Psi).
\ee
The ratio $P_i/P_0$ appearing in the redshift formula \eqref{redshift} is therefore given by
\bea
\frac{P_i(\eta_o)}{P_0(\eta_o)}=  {\tilde n}^i (1-\Phi_o-\Psi_o),~~~~~~~~~~~~~~~~~~~~~~~~~~~~\\
\label{Pi/P0}
\frac{P_i(\eta_e)}{P_0(\eta_e)}=  {\tilde n}^i (1-\Phi_e-\Psi_e)+\delta {n}^i,\qquad 
{\delta \n} = -\int _{\eta_e}^{\eta_o} d\eta \nabla_\perp (\Phi+\Psi),
\eea
where $\nabla_\perp = (\nabla- \n  \n\cdot\nabla)$, and we used the fact that, for any function $f$,
\be
\int_{\eta_e}^{\eta_o} d\eta (\d_\eta - \n \n\cdot \nabla) f(\eta,\n(\eta_o-\eta)) = f_o - f_e.
\ee
Substitution in \eqref{redshift} gives 
\be\label{To/Te}
\frac{T(\bsb{\hat n})}{T(\eta_e,x_e)}= \frac{a_e}{a_o} \exp[\Phi_e-\Phi_o +I(\eta_e,\eta_o)]
\left(\sqrt{1+v_o^2}+\bsb{\hat{\tilde n}}\cdot \bsb v_o\right)
\left(\sqrt{1+v_e^2}+(\bsb{\hat{\tilde n}}+{\delta \bsb{\hat{n}}})\cdot \bsb v_e\right)^{-1}\!\! ,
\ee
where quantities with index $e$ are evaluated at $(\eta_e,\x_e)$. In general $\eta_e$ does not have to be the same in all directions. As discussed in the text, taking the emission points to lie on a hypersurface of constant average temperature $T_{\rm rec}$ corresponds to different choices of $\eta_e$ at different directions depending on the local average of $\Phi_e$. 

In matter-domination $\Phi=\Psi = \rm{const.}$ to first order, so the ISW effect \eqref{I} starts at second order in perturbations. (Furthermore, as shown in \cite{Creminelli_SW} it is negligible in the squeezed limit, but it is essential for our study of long-long effects in section \ref{long}.) Keeping up to second order
\be
\begin{split}
\frac{T(\bsb{\hat n})}{\bar T} -1
=&\Theta_e +\Phi_e+\bsb{\hat n}\cdot (\bsb v_o-\bsb v_e)+I-{\delta \n}\cdot \bsb v_e +
\bsb{\hat n} \cdot (\bsb v_o-\bsb v_e)(\Theta_e+\Phi_e)\\
&+\Phi_e\Theta_e +\frac{1}{2}\Phi_e^2
+(\bsb{\hat n} \cdot (\bsb v_o-\bsb v_e))^2-\frac{1}{2}(\bsb v_o-\bsb v_e)^2.
\end{split}
\ee
We have defined 
\be
\Theta_e (\eta_e,x_e)= \frac{T(\eta_e,x_e)}{\expect{T(\eta_e,\x)}_\x}-1,
\ee
where the average is over the whole time-slice $\eta_e$, and 
\be
 \quad \bar T = \frac{a_e}{a_o}e^{-\Phi_o} \expect{T(\eta_e,\x)}_\x.
\ee
Finally, the coordinate of the emission point can be derived to first order by integrating \eqref{Pi/P0}, with $\eta_e$ replaced by an arbitrary time, along the line of sight, and using
\be
\int_{\eta_e}^{\eta_o}d\eta \int_{\eta}^{\eta_o} d\eta' f(\eta') = \int_{\eta_e}^{\eta_o} d\eta f(\eta).
\ee
This results in
\be
\bsb x_e = \bsb x_o + (\bsb{\hat n} -\bsb v_o + \bsb{\hat n} \bsb{\hat n}\cdot \bsb v_o)(\eta_o -\eta_e) + 2\bsb{\hat n} \int_{\eta_e}^{\eta_o} d\eta \Phi 
- 2\int_{\eta_e}^{\eta_o} d\eta (\eta-\eta_e) \nabla_\perp \Phi,
\ee
where $\bsb x_o$ is the observer's position, which we retained since it will be affected by the long gradient mode. Note that the time integrals are along the unperturbed photon trajectory $(\eta, \n(\eta_o-\eta))$.

\newpage
\section{Second order metric in matter dominated universe\label{second}}

The second order expressions for metric fluctuations in matter dominated universe read \cite{Creminelli_action}
\bea
 \Phi =  \phi+
  \left[\d^{-2}(\d_j\phi)^2
    -3\d^{-4}\d_i\d_j(\d_i\phi\d_j\phi) \right]~~~~~~~~~~~~~~~~~~~~ \label{Phi} \\
~~~~~~~~~~~~~~~~~   +\frac{1}{12}\eta^2(\d_j\phi)^2+\frac{5}{42}\eta^2\d^{-2}
\left[(\d^2\phi)^2-(\d_i\d_j\phi)^2\right] \, , \nonumber
\eea
\bea
  \Psi = \phi -\left[\frac{2}{3} \d^{-2}(\d_i\phi)^2
    - 2 \d^{-4} \d_i\d_j(\d_i\phi\d_j\phi) \right] \label{Psi}~~~~~~~~~~~~~~~~~~ \\
~~~~~~~~~~~~~~~~~   +\frac{1}{12}\eta^2(\d_j\phi)^2+\frac{5}{42}\eta^2\d^{-2}
\left[(\d^2\phi)^2-(\d_i\d_j\phi)^2\right] \, , \nonumber
\eea
\be
  \omega_{i} = - \frac{4}{3}\eta \d^{-2} \left[\d^2 \phi \d_i \phi
    - \d^{-2} \d_i\d_j
    (\d^2\phi \d_j\phi) \right] \, ,~~~~~~~~~~~~~~~~~ \label{omega} \\
\ee
\be
  {\gamma}_{ij} =  - 20
  \left( \frac{1}{3} - \frac{j_1(p\eta)}{p \eta} \right)
  \d^{-2} P_{ij \, kl}^{\rm TT}
  \left( \d_k\phi\d_l\phi \right). ~~~~~~~~~~~~~~~~~~~\label{gamma} \, 
\ee
Note that our definition of $\Phi$ and $\Psi$ in \eqref{metric} slightly differs from that of \cite{Creminelli_action}, and we have used $\eta=2/aH$ valid in a matter-dominated universe. We also used the identity
\be
\d_i \phi \d_i \d^2\phi = \frac{1}{2}\d^2(\d\phi)^2 -(\d_i\d_j\phi)^2.
\ee
The initial condition $\phi$ is proportional to the comoving-gauge scalar curvature ($\phi=-3\zeta/5$). In the expression for the tensor component $p=\sqrt{\d^2}$, the spherical Bessel function $j_1(x)$ is given by $j_1(x)=\sin(x)/x^2-\cos(x)/x$, while $P_{ij \, kl}^{\rm TT}$ is a transverse traceless projector defined as
\begin{equation}
\label{projector}
P_{ij \, kl}^{\rm TT}\equiv \frac12 \left(P_{ik} P_{jl} + P_{jk} P_{il}- P_{ij} P_{kl}\right) \;,
\end{equation}
where $P_{ij}$ is a symmetric transverse projector given by
\begin{equation}
P_{ij}\equiv \delta_{ij} -{\partial_i\partial_j\over \partial^2} \;.
\end{equation}
$P_{ij \, kl}$ can be expanded to give
\begin{equation}\label{expansion2}
  P_{ij \, kl}^{\rm TT} \left( \d_k\phi\d_l\phi \right) = - \d^{-2}
  \left[ \d^2 \Omega \delta_{ij} + \d_i \d_j \Omega +2 (\d^2 \phi
    \d_i \d_j \phi - \d_i \d_k \phi \d_j \d_k \phi) \right]\;,
\end{equation}
with
\begin{equation}
  \Omega = - \frac12 \d^{-2} \left[ (\d^2 \phi)^2 -(\d_i\d_j\phi)^2
  \right] \, .
\end{equation}

\newpage
\section{Locality of physical observables\label{local}}
\subsection{quadrupole and higher multipoles\label{multi}}

Neglecting $\eta_e/\eta_o$, and focusing on the contribution to quadrupole and higher multipoles, the time-dependent pieces matter only in the ISW term and the time-independent non-local piece in the linear SW: $\Phi/3$. To show that the non-localities cancel, it is sufficient to expand $\gamma_{ij}$ to order $\eta^4$. Working in momentum space, and after a few partial integrations, different contributions to the ISW term can be brought into the form
\be
\begin{split}
I_{\Phi+\Psi}=&\int d^3\bsb p_1 d^3\bsb p_2 \phi_{\bsb p_1}\phi_{\bsb p_2} \int d\eta \eta\left[\frac{10}{21}
\frac{(\bsb p_1\cdot\bsb p_2)^2- p_1^2 p_2^2}{p^2} -\frac{1}{3}\bsb p_1\cdot\bsb p_2\right]e^{i\bsb p\cdot \bsb{\hat n} (\eta_o-\eta)}\nonumber\\[10pt]
I_\omega =&\int d^3\bsb p_1 d^3\bsb p_2 \phi_{\bsb p_1}\phi_{\bsb p_2}\left[\int d\eta \frac{4}{3}\eta (\bsb p\cdot \bsb{\hat n})
\frac{p_1^2 \bsb p_2\cdot \bsb{\hat n}}{p^2} e^{i\bsb p\cdot \bsb{\hat n} (\eta_o-\eta)}\right.\\[10pt]
&~~~~~~~~~~~~~~~~~~~~~~~~~~~~~~~~~
\left.+\frac{4}{3}\left(e^{i\bsb p\cdot \bsb D}-1\right)\frac{p_1^2\bsb p\cdot \bsb p_2}{p^4}
-\frac{4}{3}i\eta_o\frac{p_1^2\bsb p_2\cdot \bsb{\hat n}}{p^2}\right],\nonumber\\[10pt]
I_\gamma =& \int d^3\bsb p_1 d^3\bsb p_2 \phi_{\bsb p_1}\phi_{\bsb p_2}\left\{\int d\eta \eta \left[
\frac{10}{21}\frac{p_1^2 p_2^2-(\bsb p_1\cdot\bsb p_2)^2}{p^2} -\frac{4}{3}\frac{p_1^2\bsb (\p_2\cdot\bsb{\hat n})^2}{p^2}  +\frac{4}{3}\frac{\bsb p_1\cdot\bsb p_2 \bsb p_1\cdot\bsb{\hat n} \bsb p_2\cdot\bsb{\hat n}}{p^2}\right]e^{i\bsb p\cdot \bsb{\hat n} (\eta_o-\eta)}\right.\nonumber\\[10pt]
&~~~~~~~~~~~~~~~~~~\left.-\frac{1}{3}\left[\left(e^{i\bsb p\cdot \bsb D}-1 - i \bsb p\cdot \bsb{\hat n}\eta_o\right)
+\frac{1}{14}p^2\eta_o^2(3+ i \bsb p\cdot \bsb{\hat n}\eta_o)\right]\frac{p_1^2 p_2^2-(\bsb p_1\cdot\bsb p_2)^2}{p^4}\right\},\nonumber
\end{split}
\ee
where $\p = \p_1+\p_2$. Summing them up the non-localities in the integrals cancel against one another, and when included in the Sachs-Wolfe formula, the boundary terms cancel the time-independent non-local pieces in $\Phi$ up to a monopole and a dipole:
\be
\frac{1}{3}\Phi+I= \frac{1}{3}\phi 
-\frac{1}{3}\int d\eta \eta [2(\bsb{\hat n} \cdot \nabla \phi)^2-(\d_j \phi)^2]+\Theta_0+\Theta_1 + \cdots
\ee
where dots correspond to local, higher derivative terms from the expansion of $j_1(p\eta)$ in $\gamma_{ij}$. The non-local expressions for monopole and dipole are
\be
\label{Theta0}
\begin{split}
\Theta_0=&\left\{\frac{1}{3}\frac{1}{\d^2}\left[(\d \phi)^2-3\frac{1}{\d^2}\d_i\d_j(\d_i\phi\d_j\phi)\right]+\frac{1}{14}\eta_o^2\frac{1}{\d^2}\left[(\d^2\phi)^2-(\d_i\d_j\phi)^2\right]\right\}_{\bsb x =0}\\[10pt]
\Theta_1=&\left\{\frac{1}{3}\eta_o \left(1+\frac{1}{14}\eta_o^2\d^2\right)\bsb{\hat n}\cdot \nabla\frac{1}{\d^4}
\left[(\d^2\phi)^2-(\d_i\d_j\phi)^2\right]-\frac{4}{3}\eta_o \frac{1}{\d^2}(\d^2\phi \bsb{\hat n}\cdot \nabla \phi)\right\}_{\bsb x =0}.
\end{split}
\ee

\subsection{Monopole and dipole\label{mondi}}

As already seen in the analysis of the long-short effect the naive expressions for monopole and dipole are degenerate with the observer's time $\eta_o$ and velocity $\v_o$. Here we show that the above ambiguous results cancel in the physically well-defined quantities. The part of the solution \eqref{Phi}-\eqref{gamma} which is responsible for these non-local expressions is locally equivalent to a gauge transformation. Therefore, as in section \ref{LS} the change of $T(\n)$ due to the variation of observer's $\Phi_o,\eta_o,$ and $\v_o$ cancel against the naive contribution to monopole and dipole.

Let us parameterize the non-local part of the solution which is relevant for monopole and dipole as follows
\be
\label{Phi2}
\Phi = a_1 + \bsb b_1 \cdot \bsb x + \eta^2(a_2+\bsb b_2 \cdot \bsb x)
\ee
where
\be
a_{1,2} = f_{1,2}(\bsb 0),\qquad \bsb b_{1,2} = \nabla f_{1,2}(\bsb 0),
\ee
and 
\bea
f_1 &=& \left[\d^{-2}(\d_j\phi)^2 -3\d^{-4}\d_i\d_j(\d_i\phi\d_j\phi) \right]\\[10pt]
f_2 &=& \frac{5}{42}\eta^2\d^{-2}\left[(\d^2\phi)^2-(\d_i\d_j\phi)^2\right].
\eea
We also have for the shift vector 
\be\label{omega2}
  \omega_{i} = - \frac{4}{3}\eta \d^{-2} \left[\d^2 \phi \d_i \phi
    - \d^{-2} \d_i\d_j
    (\d^2\phi \d_j\phi) \right]_{\x=\bsb 0} \, .~~~~~~~~~~~~~~~~~ 
\ee
These metric perturbations can be generated by the following coordinate transformation
\be\label{trans}
\begin{split}
\eta\; \to\; 
& \eta + \frac{1}{3}\eta (a_1 + \bsb b_1 \cdot \bsb x) +\frac{1}{5} \eta^3(a_2+\bsb b_2 \cdot \bsb x),\\[10pt]
\x \;\to\; & x + \frac{1}{2} \eta^2 \bsb b_3 + \frac{1}{4}\eta^4\bsb b_4 
\end{split}
\ee
if $\bsb b_{3,4}$ are chosen such that
\be
\bsb{\omega} = \eta(\bsb b_3 -\frac{1}{3}\bsb b_1) +\eta^3 (\bsb b_4 -\frac{1}{5} \bsb b_2),
\ee
agrees with \eqref{omega2}. Using the identity
\be
\d_i\d_j(\d_i\phi\d_j\phi)-\d^2(\d_j\phi)^2 = (\d^2\phi)^2-(\d_i\d_j\phi)^2,
\ee
we find
\be
b_3^i =\frac{1}{3}\frac{\d_i}{\d^4}[(\d^2\phi)^2-(\d_i\d_j\phi)^2]-\frac{4}{3} \frac{1}{\d^2}(\d^2\phi \d_i \phi),
\ee
and 
\be
b_4^i = \frac{1}{42}\frac{\d_i}{\d^2}[(\d^2\phi)^2-(\d_i\d_j\phi)^2].
\ee
This transformation generates the following expression for $\Psi$
\be
\label{Psib}
\Psi = -\frac{2}{3}(a_1 + \bsb b_1 \cdot \bsb x) - \frac{2}{5}\eta^2(a_2+\bsb b_2 \cdot \bsb x).
\ee
The contribution of this gauge mode to the monopole and dipole comes from the linear plus integrated Sachs-Wolfe $\Phi/3 + I$. (We work in the approximation $\eta_{\rm rec}=0$.) The result is
\be
\begin{split}
\Theta_0 \;=\;& \frac{1}{3}a_1+\frac{3}{5} a_2 \eta_o^2\\[10pt]
\Theta_1 \;=\;& \eta_o \bsb b_3 \cdot \n + \eta_o^3 \bsb b_4\cdot \n,
\end{split}
\ee
which coincide with the non-local expressions (\ref{Theta0}). The non-local contribution of the remaining part of $\Psi$, i.e. the difference between \eqref{Psi} and \eqref{Psib}, and $\gamma$ in \eqref{gamma} cancel against each other.

To see the cancellation of $\Theta_0$ in the physical observables, note that the observer's proper time is
\be
t_* = \int^{\eta_o}_0 d\eta \eta^2 (1+\Phi) = \frac{1}{3}\eta_o^3(1+a_1 + \frac{3}{5} a_2 \eta_o^2).
\ee
So to measure CMB temperature at the same proper time we should shift
\be
\label{eta_o}
\eta_o = \eta_* (1 - \frac{1}{3} a_1 -\frac{1}{5}a_2\eta_o^2).
\ee
The observed average temperature is
\be
\expect{T_o(\n)} = e^{-\Phi_o} \frac{a_e}{a_o}\expect{T(\eta_e,\x_e)}_{\x_e}.
\ee
In the presence of a gravitational potential 
\be
\label{aeT}
a_e\expect{T(\eta_e,\x_e)}_{\x_e} = a_{\rm rec} T_{\rm rec} (1+\frac{1}{3}\Phi_e + I)=a_{\rm rec} T_{\rm rec}(1+\Theta_0).
\ee
On the other hand, using \eqref{Phi2} and \eqref{eta_o}
\be
e^{\Phi_o}a_o = e^{\tilde \Phi_o} a_* (1+\frac{1}{3}a_1 + \frac{3}{5}a_2\eta_o^2)
\ee
which cancels $\Theta_0$ in \eqref{aeT}. Finally, the transformation \eqref{trans} shifts the observer's velocity to
\be
\v_o = \tilde\v_o - \eta_o \bsb b_3 + \eta_o^3 \bsb b_4,
\ee
which cancels $\Theta_1$.


\end{document}